\documentclass{article}

\usepackage[final]{neurips_2023}

\usepackage[utf8]{inputenc} 
\usepackage[T1]{fontenc}    
\usepackage{hyperref}       
\usepackage{url}            
\usepackage{booktabs}       
\usepackage{amsfonts}       
\usepackage{nicefrac}       
\usepackage{microtype}      
\usepackage{xcolor}         
\usepackage{mathrsfs,amsmath}
\usepackage{graphicx}

\title{Neural Architectures Learning Fourier Transforms, Signal Processing and Much More....}

\author{%
 Prateek Verma\\
 Stanford University\\
  Stanford, CA, USA 94305 \\
  \texttt{prateekv@stanford.edu} \\
}

\begin{document}

\maketitle

\begin{abstract}

This report will explore and answer fundamental questions about taking Fourier Transforms and tying it with recent advances in AI and neural architecture. One interpretation of the Fourier Transform is decomposing a signal into its constituent components by projecting them onto complex exponentials. Variants exist, such as discrete cosine transform that does not operate on the complex domain and projects an input signal to only cosine functions oscillating at different frequencies. However, this is a fundamental limitation, and it needs to be more suboptimal. The first one is that all kernels are sinusoidal: What if we could have some kernels adapted or learned according to the problem? What if we can use neural architectures for this?  We show how one can learn these kernels from scratch for audio signal processing applications. We find that the neural architecture not only learns sinusoidal kernel shapes but discovers all kinds of incredible signal-processing properties. E.g., windowing functions, onset detectors, high pass filters, low pass filters, modulations, etc.
Further, upon analysis of the filters, we find that the neural architecture has a comb filter-like structure on top of the learned kernels. Comb filters that allow harmonic frequencies to pass through are one of the core building blocks/types of filters similar to high-pass, low-pass, and band-pass filters of various traditional signal processing algorithms. Further, we can also use the convolution operation with a signal to be learned from scratch, and we will explore papers in the literature that uses this with that robust Transformer architectures. Further, we would also explore making the learned kernel's content adaptive, i.e., learning different kernels for different inputs.

\end{abstract}

\section{Introduction and Motivation}
For any continuous signal $f(t)$ with period 1, the Fourier series is a way of decomposing any signal into a sum of sinusoidal functions, namely,

$$
f(t)=\sum_{k=1}^N a_k \sin \left(2 \pi k t+\phi_k\right) \\
$$

where $a_k$ is the weight associated with the $k^{th}$ sinusoidal basis and $\phi_k$ is the phase or the time shift associated with it. Rearranging the terms, we can now associate the same expression in terms of a simplistic sum of complex exponentials, which is,

$$f(t)=\sum_{k=-n}^n c_k e^{2 \pi i k t}=\sum_{k=-n}^n\left\langle f, e^{2 \pi j k t}\right\rangle e^{2 \pi j k t}$$

where now $c_k$ becomes complex coefficient, and the basis functions $e^{2 \pi i k t}$ are themselves complex. As seen in \cite{osgood2002lecture}, one can interpret, the coefficient $c_k$ themselves as projections of a function $f$ onto the complex exponential basis, where  

$$
c_m=\int_0^1 f(t) e^{-2 \pi i m t} d t 
$$

and we define the inner product as, 

$$
\left\langle f, e^{2 \pi j k t}\right\rangle=\int_0^1 f(t) \cdot e^{-2 \pi j k t} d t
$$

The same expressions in any signal can be adapted by making the signal period $T$, making $T$ tending to $\infty$. The same expressions now for signals with arbitrary period $T$ become,

$$c_k=\frac{1}{T} \int_{-T / 2}^{T / 2} e^{-2 \pi j k t / T} f(t) d t $$.

Now, a function $f$ with period $T$ can be expressed in terms of these coefficients as,

$$f(t)=\sum_{k=-\infty}^{\infty} c_k e^{2 \pi j k t / T} $$

These expressions can now be written as, 

$$\mathcal{F}f(s)=\int_{-\infty}^{\infty} e^{-2 \pi j s t} f(t) d t  \quad f(t)=\int_{-\infty}^{\infty} \mathcal{F}f(s) e^{2 \pi j s t} d s 
$$

by defining $s = k/T $. There exist similar expressions for discrete signals giving rise to Discrete Fourier Transform. For computers dealing with real-world signals, e.g., audio, we cannot process continuous signals and has to be stored in digital format. For this scenario, we must devise a way to convert continuous time domain signals to discrete digital signals. This is carried out in three steps, and for detail, the reader can refer to \cite{osgood2002lecture}. 
The first step is finding a reasonable discrete approximation to $f(t)$, namely $f_d(t)$, and using similar techniques to find a discrete approximation to $\mathcal{F}f_d(s)$, i.e., $\mathcal{F_S}f_{d}(s)$, which we called the sampled version of continuous Fourier Transform. Final step includes a way of passing and connecting $f_d(t)$ to  $\mathcal{F_D}f_{d}(s)$. This is the path is again explained in \cite{osgood2002lecture}, but for the sake of brevity, we define them here:

$$
\begin{aligned}
& \underline{F}  \underline{f}[m]=\sum_{n=0}^{N-1} \underline{f}[n] e^{-2 \pi i n m / N} \\
& \underline{F}=(\underline{F}[0], \underline{F}[1], \underline{F}[2] \ldots . .)_{[0, M-1]} \\
& \underline{f}=(\underline{f}[0], \underline{f}[1], \underline{f}[2] \ldots)_{[0, N-1]}
\end{aligned}
$$

where $\underline{F}$ is a $M$ length sequence corresponding to the sampled version of taking a discrete-time Fourier Transform and $\underline{f}$ is a $N$ length sampled version for a continuous-time signal $f(t)$. We can see stark similarities between this with the continuous version. 

A variant of the Discrete Fourier Transform is a Discrete Cosine Transform (DCT) which deals with real signals getting mapped to real sinusoidal basis functions \cite{ahmed1974discrete}. A DCT can express the same rudimentary idea but can now thus express a finite sequence of data points as a finite/infinite sum of cosine functions. A discrete cosine transform (DCT) expresses a finite sequence of data points in terms of a sum of cosine functions oscillating at different frequencies and is defined as,

$$
X_k=\frac{1}{2}\left(x_0+(-1)^k x_{N-1}\right)+\sum_{n=1}^{N-2} x_n \cos \left[\frac{\pi}{N-1} n k\right] \quad \text { for } k=0, \ldots N-1
$$

where $x_n$ is a length of N numbers getting (discrete cosine) transformed to a sequence of length $N$ numbers. The goal of this work is again to go from N-length real sequences to N-length real sequences, but using kernels instead of cosine functions learned from data via neural architectures. Further, we will make these kernels' content adaptive, i.e., depending on the input, we will have different kernels, thus making the transform content adaptive \cite{}. 

In order to motivate this work, we set at the outset the main goals. For the rest of this work, we will deal with 1-D signals, which are audio signals, sounds that we hear every day. If we glance over the expressions above, we find that the primary expression is sub-optimal in several ways, namely, \newline.

i) \textbf{Type of signal}: What if we have only a signal of a specific category present in our problem/dataset? To give a  very coarse example, let us say we only deal with square waves for the problem of interest. In this scenario, it would make sense to have the fundamental building blocks as square waves rather than pure sinusoids, even if they do not satisfy the orthogonality property. 

ii) \textbf{Learning the signal}: It would be even better if we could have different functions that can be learned according to the dataset, which can be different according to different problems/data of interest.

iii) \textbf{Frequency of the basis functions}: We see another sub-optimality in the expression above, i.e., all of the frequencies of the sinusoidal basis are linearly spaced. We say we only have frequency content in a particular range or bands of frequencies; it is optimal to have the distribution of the basis functions similar to that in the input.

iv) \textbf{Adaptive}: Can we make the basis functions or the kernels that are learned themselves adaptable and learnable? So let us say for a given dataset of 1-D signals, it is a mixture of music and speech. Can we learn a set of basis functions for music and the other set for speech, and switch the basis function depending on the content of th e input signals.

\section{Preliminaries} We will, for the sake of completeness, describe what a short-time Fourier Transform is, how it is different from taking a DFT, and how it is used to represent signals that are changing across time.

\subsection{Neural Networks}
A Neural Architecture is designing a function through a series of matrix operations onto a vector or a matrix to give another vector or matrix output. The simplest architecture can be considered a chain of linear matrix operations cascaded by a non-linear operation, typically a relu operator. For example, a 3-layer fully connected MLP architecture is, 

$$y=W_3^{\top}\left(\sigma\left[W_2^{\top}\left(\sigma\left[w_1^{\top} x+b_1\right]\right)+b_2\right]\right)+b_3$$

with the input being $x$, and $\sigma$ being the vector non-linearity which is $max(0,x)$. They usually are trained with the weights initialized with random weights and updated by the backpropagation algorithm to minimize a certain objective function. Even complex formulations like the latest work on Transformers \cite{vaswani2017attention}, that can learn complex connections for autoregressive architectures \cite{verma2021generative} follow similar steps; a network is initialized with random weights and is updated to minimize an objective criterion. For problems such as classification, we typically minimize the cross-entropy loss. The target label (e.g., if an image is a cat/dog) is converted to one-hot encoding, where the total number of choices is converted to the index where the desired label lies (e.g., the cat is given scalar 0, dog one and so on and putting "one-hot "1 at the appropriate location). Now the output of the neural architecture is converted to a probabilistic output via a softmax operation, and we minimize the output of the prediction of the network with that of the target one-hot label via cross-entropy loss. The cross-entropy loss $L$ is defined as,

$$\mathcal{L}=-\sum_{i=1}^{\text {output-size }} y_i \log \hat{y}_i$$

where $\hat{y}_i$ is the predicted output and $y_i$ is the target output. For continuous-valued regression problems, or in cases to predict multiple outputs, we typically use a Huber Loss \cite{huber1992robust}  between the prediction $f(x)$ and the target labels $y$, as 

$$
L_\delta(y, f(x))= \begin{cases}\frac{1}{2}(y-f(x))^2 & \text { for }|y-f(x)| \leq \delta \\ \delta \cdot\left(|y-f(x)|-\frac{1}{2} \delta\right), & \text { otherwise }\end{cases}$$

A typical advancement in a neural architecture either occurs with the use of a complex optimization function \cite{verma2018neural}, or with using the same optimization function but advancing the architecture itself \cite{verma2020framework}, or sometimes using multiple loss functions together in a multi-criterion setting \cite{guo2019end}.

\subsection{Short Time Fourier Transform}
Typically, a Short-Time Fourier Transform  (STFT) is defined as taking windowed patches of a continuous signal and taking DFT on those patches. It is defined as,

$$X_m\left(\omega_k\right)=\sum_{n=-\infty}^{\infty}[w(n-m) x(n)] e^{-j \omega_k n}$$

where $w$ is a windowed signal which is non-zero on a finite interval, and the rest of the definition is similar to taking a DFT on this modified signal \cite{allen1977unified}. An STFT can also be interpreted as the output of a filter bank where the terms can be rearranged to give the following output. The original expression can be re-written as, 

$$
\begin{aligned}
X_m\left(\omega_k\right) & =\sum_{n=-\infty}^{\infty} \underbrace{\left[x(n) e^{-j \omega_k n}\right]}_{x_k(n)} w(n-m) \\
& =\left[x_k * \operatorname{FLIP}(w)\right](m)
\end{aligned}
$$
Now, "STFT is thus interpreted as a frequency-ordered collection of narrow-band time-domain signals, as depicted in Fig.9.2. In other words, the STFT can be seen as a uniform filter bank in which the input signal $ x(n)$ is converted to a set of $ N$ time-domain output signals  $ X_n(\omega_k)$,  $ k=0,1,\ldots, N-1$, one for each channel of the $ N$ -channel filter bank" \cite{SASPWEB2011}. This definition will help us later explore what happens when we have a bank of filter banks and choose one of the possible $ N $ outputs of the filter banks.

\begin{figure*}[t]
  \centering
  \hspace*{8.8pt}
  \includegraphics[width=\linewidth, height=5cm]{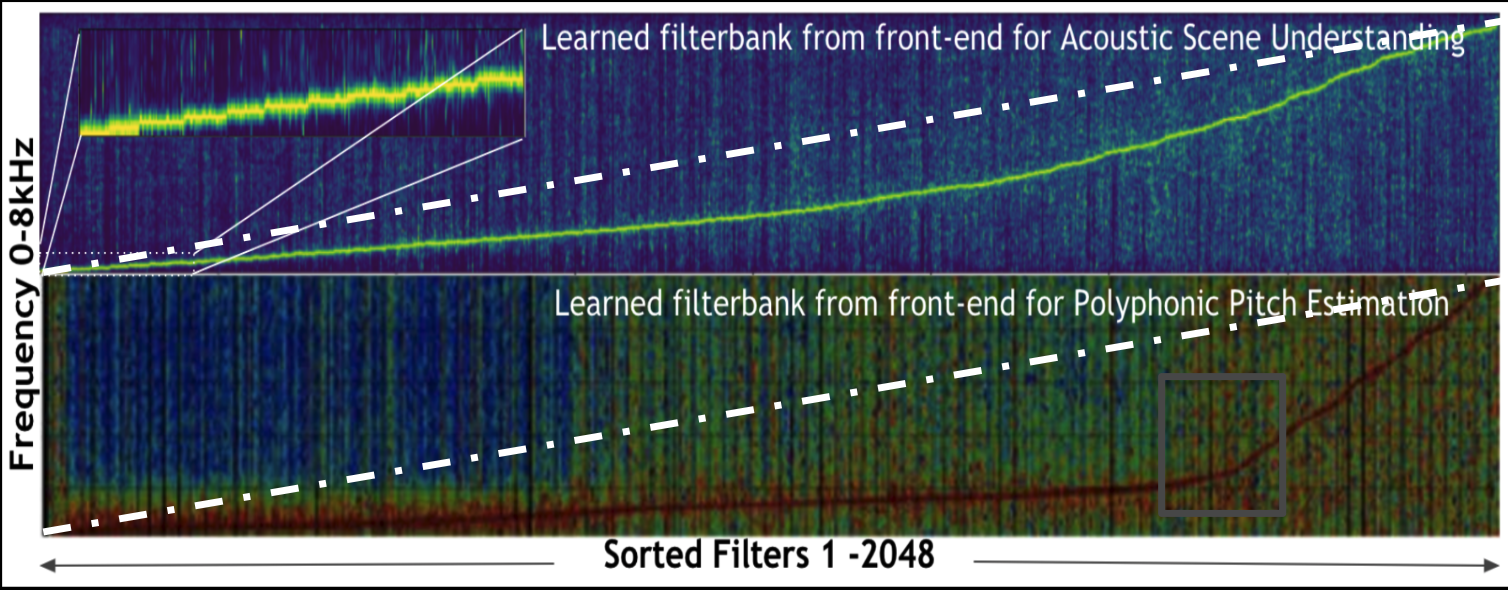}
  \caption{We plot the frequency response of a single layer of 2048 neurons for two different tasks: acoustic scene understanding as well as pitch estimation. We see that the response varies according to that of task of interest. In addition it also is quite different than a traditional Discrete Fourier Transform}
  \label{fig:speech_production}
\end{figure*}

\section{Neural architectures learning signal processing}
We, in this section will show how neural architectures can learn core signal processing ideas such as non-linear non-constant bandwidth filter-bank, comb filters, windowing functions, and first-order difference functions. Further, we will show how one can approximate a slice of a spectrogram with that of a single convolution filter followed by max-pooling and a log operation. Finally, we will also explore how we can learn the front-end or a learned spectral representation that is adaptive according to the task and the input signal itself. This is akin to making Fourier Transforms content adaptive.

\subsection{Learning Time-Frequency Representation Kernels}
We train a neural architecture with two layers with the following setup. The input given is a waveform of 40ms or 640 samples at 16KHz. The MLP layers are nothing but a cascade of $W^Tx+b$ followed by relu non-linearities. This is similar to the expression of a DFT, but now each of the filters or the basis functions/kernels are learned from scratch by training a neural architecture via an optimal loss function. 

$$
\begin{aligned}
\underline{\mathcal{{F}}} \underline{f}[m] & =\sum_{n=0}^{N-1} f[n] e^{-2 \pi i m / n} \\
& =\sum_{n=0}^{N-1} f[n] \Psi_{learned}[m, n]
\end{aligned}
$$
In our case, we trained the architecture to find a signal input's fundamental frequency in $x[n]$. If the signal contains noise or is not a periodic signal, then we assign a separate flag to classify it; otherwise, the output is a discrete index ranging from 0-77 to cover a frequency range from 40Hz to 540 Hz, with each of the indices in the log scale (cent scale). This mimics how humans hear. We pass on about a million snippets of a 40ms waveform and the output discrete label. We try to minimize the cross-entropy loss between the predicted output and the one-hot encoding of the ground truth labels. The weights are updated continuously using backpropagation algorithm\cite{lillicrap2020backpropagation}, using Tensorflow framework \cite{abadi2016tensorflow}. Once trained, we start looking at each of the neurons that are nothing but $\Psi_{learned}[m, n]$, with $m$ ranging from 0 to 2048. These kernels are patches of the waveform that is multiplied or, in other words, the output of taking a dot product with that of the kernel to produce a single scalar output. An important distinction from DFT is that these neurons would all be shuffled up. We now sort these filters according to how the properties they exhibit. One of the properties is to sort them according to the frequency content of these neurons. How? We take the waveform/weight that each of the neurons learns, which is nothing but a vector of dimension the same as that of input of length $N$. We now compute the DFT and sort all of these 2048 neurons according to the location of the peak. We then stack all sorted neurons, as shown in the Figure 1. We find that, as seen from the Figure 1, it learns characteristics different from a traditional DFT. If it were to learn a DFT, the sorted peaks would be a straight line connecting bottom left to top right. We see that the neurons learn a non-linear step-wise frequency response. It assigns a lot more filters to the lower frequencies and only a few filters or basis functions outside of the range of human voice. Secondly, it also assigns different bandwidths to filters that get fired when a particular frequency is present. This can be seen from the fact that in the Figure 1, different color contrasts are present. The initial filters are low bandwidth and only allow selective filters to pass through. In contrast, the filters that allow a more comprehensive range of frequencies are present when the center frequency is higher.

\begin{figure*}[t]
  \centering
  \hspace*{8.8pt}
  \includegraphics[width=\linewidth, height=5cm]{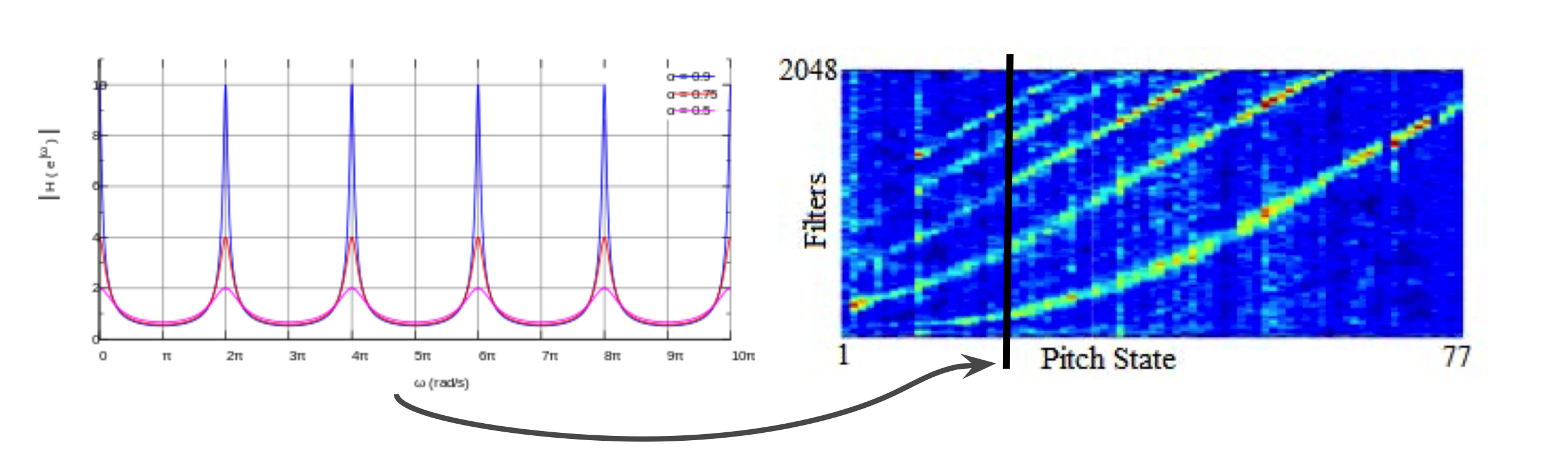}
  \caption{We plot the frequency response of the comb filter on the left taken from and on the right the discovered comb filter bank. Notice how for each pitch state akin to a traditional comb filter, the neuron or the filter weight learns to sum up the locations similar to the comb when we sort the filters in the previous layer }
  \label{fig:speech_production}
\end{figure*} 
\subsection{Discovering Comb filters}

In the past subsection, we saw what each neuron is learning. In this work, we try to understand what potentially a second layer of a neural architecture would do and tie it together with the core idea in discrete signal processing and filtering: comb filtering. What are comb filters? Mathematically, the response $y[n]$ for a discrete filter input $x[n]$ is defined as, 

$$y[n]= x[n] + \alpha x [n-K]$$

where K is the delay, the frequency response of a comb filter consists of a series of regularly spaced notches between regularly spaced peaks (sometimes called teeth), giving the appearance of a comb. This is because of causing constructive and destructive interference in the time domain \cite{pei1998comb}. Such a filter is ideal in dealing with audio signals, as almost all speech/music, harmonic signals, consist of strength or peaks appearing in the spectrum of the signal at locations $f_o$, $2*f_o$, $3*f_o$ and so on, when we compute STFT, which is a DFT computed on windowed signal at smaller time-resolution, instead of the entire signal typically 25ms.

Now that we have set up the preliminaries, we now look at the activations of the first layer and sort them according to the peak where their frequency response exists. This would index the frequency response of each of the neurons, similar to how a traditional DFT basis would be indexed. Given this setup, adding only one more layer to this, the neuron corresponding to the pitch activation should be looking at the individual neuron of the previous layer at locations $f_o$, $2*f_o$, $3*f_o$ so that it can sum them all activations linearly, as each neuron is carrying out $W^Tx+b$. The additional thing in our finding is that since the frequencies in the first layer are not equally spread, the comb pattern should also follow a log scale. This is indeed the case, as seen in the Figure 2 for experiments on pitch detection for polyphonic music. On the left, we plot the frequency response of the comb filter. We can see spikes, which means that those frequencies can pass through occurring at harmonic locations. On the right, for a 2-layer neural architecture, each neuron in the second layer would correspond to a discrete pitch state within a specific interval $[f_1,f_2]$. Since humans hear sounds in a logarithmic scale, we split the interval in the log space for deciding $[f_1,f_2]$, one of which is the cent scale. Thus, for each pitch state neuron, it should ideally start to look at the harmonics, as the signal for speech and music exhibits that property. The Figure 2 shows that it precisely does that: we stack up all the neurons along the x-axis, which are nothing but weights assigned to our individual neuron in layer 1, and sorted filters along the y-axis. We see that it learns to sum up precisely at the location in a harmonic manner for all of the pitch states in our dataset. Surprisingly for high-pitch stats, our frequency responses in the first layer only have assigned many neurons to the range of frequencies corresponding to human speech.

\subsection{Learning time-frequency representation according to the task}
As we saw in the previous exploration, one can see that a single layer of 2048 neurons is equivalent to learning kernels onto which the contents of the input signal $x[n]$ are projected.

$$
\begin{aligned}
\underline{\mathcal{{F}}} \underline{f}[m] & =\sum_{n=0}^{N-1} f[n] e^{-2 \pi i m / n} \\
& =\sum_{n=0}^{N-1} f[n] \Psi_{task}[m, n]
\end{aligned}
$$

\begin{figure*}[t]
  \centering
  \hspace*{8.8pt}
  \includegraphics[width=0.7\linewidth]{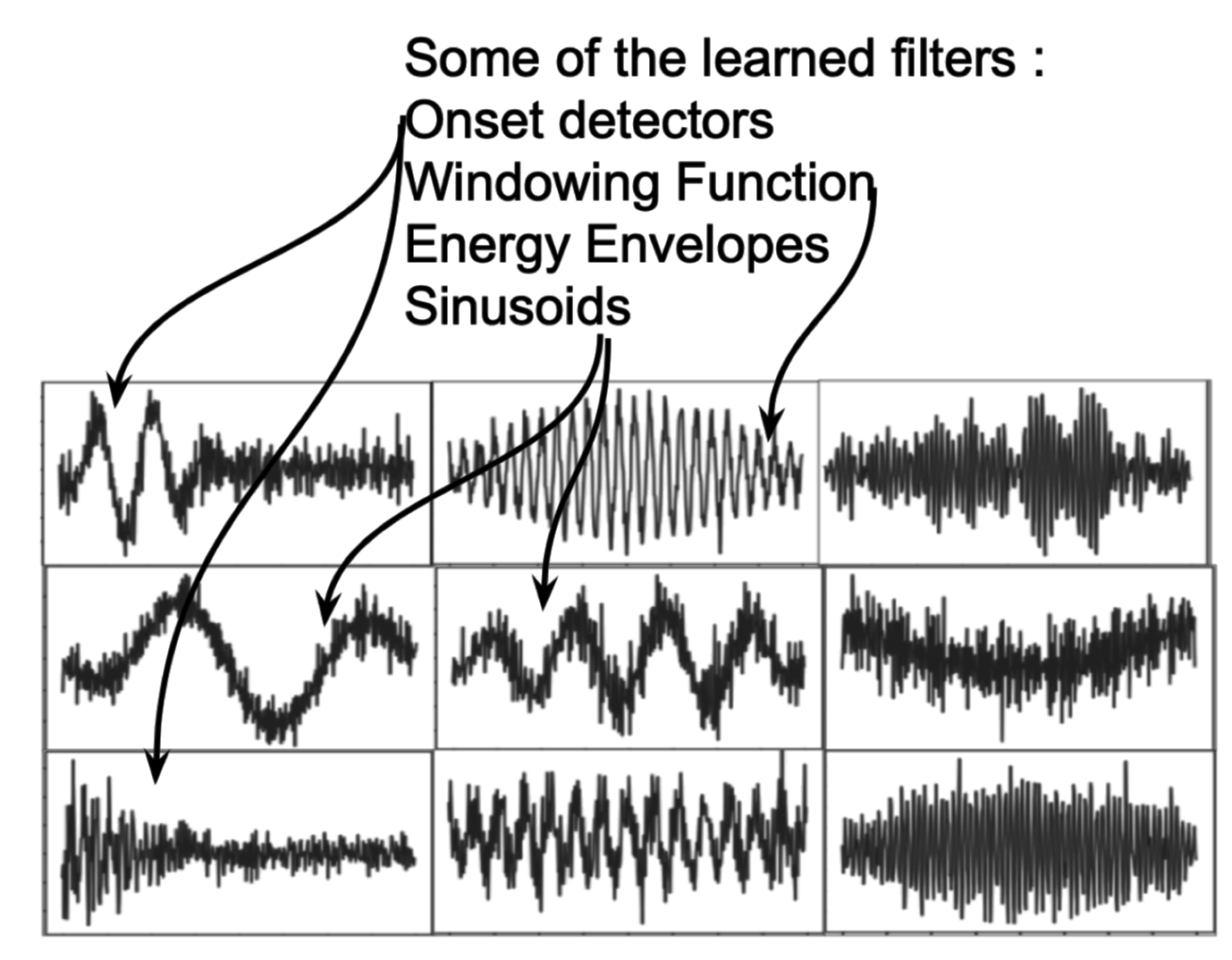}
  \caption{We compute the raw shape of the neurons which exhibit traditional signal processing properties like onset detection, windowing functions, modulations, and onset detectors, and many more.}
  \label{fig:speech_production}
\end{figure*}

If we compare what individual neurons do, we can see stark similarities of the expression $\Psi_{task}[m, n]$ with that of taking the dot product $W_m^Tx+b$. The projection of a signal $x[n]$ onto the $m^{th}$ kernel of sequence length $n$ is equivalent to $W_m^Tx+b$. With this analogy established, we can now say that instead of $\Psi[m, n]$ being fixed for \textit{every task}, we will explore if the learnable kernels learn different properties for different tasks. With training neural architecture, which is similar for two different tasks, we find that the neural architectures learn a different front end for different tasks. We train the neural architecture for two orthogonal tasks - for extracting pitch or the fundamental frequency of a given signal $x[n]$, and in the second case, given a signal $x[n]$, what would be the timbre or the contents of the input signal from a list of 200 categories. We store the weights described in the previous section upon training similar architecture for these two tasks. We now sort them according to the peak location of taking the Fourier Transform so that, similar to the exponential basis, we can sort them in increasing frequency of the kernels. We have taken this Figure 3 from \cite{verma2021audio} for reference. To our amazement, we learn different characteristics for different tasks, as can be seen in Figure 1 ! We see that both of these plots are different from a sorted purely sinusoidal kernel that is depicted by a white dashed line. For the polyphonic pitch estimation of the human voice, we see that it assigns many neurons to the lower frequencies instead of acoustic scene understanding. Further, the bandwidth assigned for each task is also different, i.e., the peaks are more prominent in lower frequencies (as depicted by high contrast or more blue present) than in the higher frequencies. This depicts that, unlike a traditional kernel of the DFT, the kernels learn non-constant bandwidth kernels.

\subsection{What do individual neurons look like: }

This section depicts the shape of each of the kernels $\Psi[m, n]$, with m being the index of the kernel and n being ranging from $[0-400]$, for each of the individual neurons (which we can also call filters). Each neuron is the kernel onto which the signal $x[n]$ is projected. From Figure 3, we handpick nine neurons and illustrate their properties. We find that the neuron learns all kinds of incredible signal-processing properties. For example, the neuron on the top left and bottom left depict onset detectors. These may correspond to sudden impulse-like events like fricatives for speech, a gunshot, or a bang, knock on the door.
Interestingly, it learns different types of onset detectors for, e.g., different decaying properties for different types of onsets. It learns a sinusoidal kernel, as seen in the center top kernel. Surprisingly, this kernel is not flat and is multiplied by a window function like a Hanning or a Hamming function, which was learned from scratch. Finally, it is learning modulations on top of the learned sinusoids. The only drawback of this approach is that it is learning a kernel that can multiply with that of the input. This, with a finite number of kernels or neurons, let us say 2048 of them, cannot capture all the phase variations present in the signal. That opens up the possibility of learning convolutional kernels to convolve the input signal with a learned convolutional filter. In the next section, we will motivate convolutions, STFT, and how we can migrate to a compelling idea in signal processing: learning a content adaptive kernel, i.e., we can pick up a kernel depending on the contents of the input signal.

\begin{figure*}[t]
  \centering
  \hspace*{8.8pt}
  \includegraphics[width=\linewidth]{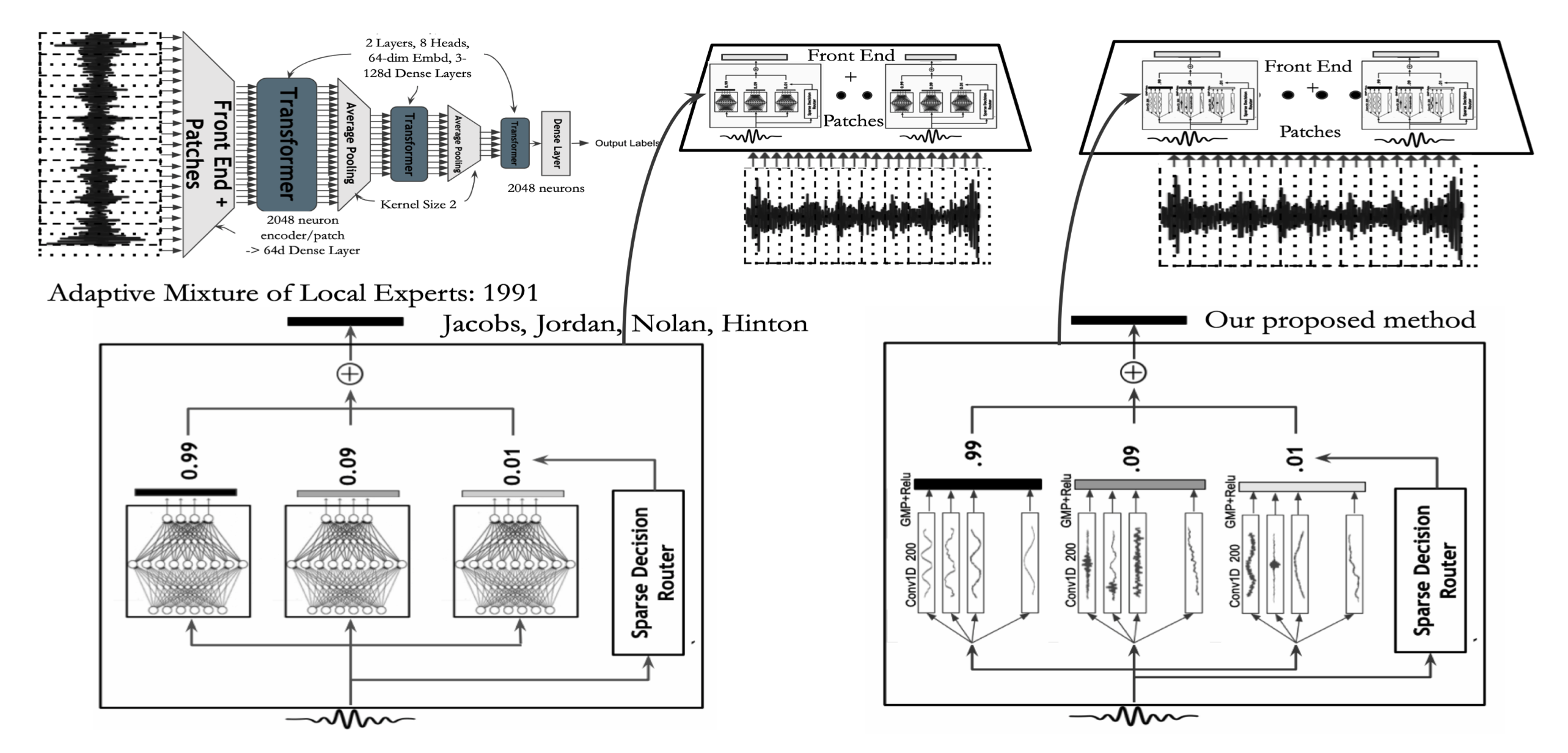}
  \caption{We show how we can learn a content adaptive Fourier like representation where depending on the contents of the input signal, we compute different representations for the input signal depending on the contents of the input signal. On the left we show a bank of kernels being learned by MLPs and on right via convolutional architecture  }
  \label{fig:speech_production}
\end{figure*} 

\section{A Content Adaptive Fourier Transform: Learning $\Psi_{input}[m, n]$}

We saw from previous sections two main takeaways: i) A STFT can have a filter-bank interpretation, and ii) For different tasks, we learn different kernels depending on the task at hand. We want to explore this further: Can we learn different kernels depending on the input signals? It makes sense to a certain extent: If our input signal contains, say, for the sake of example, a square wave, then in our list of kernels, it will be optimal to have kernels that are square rather than many sinusoids that are competing to have the closest approximation to that of square waves. Another analogy is the domain: if we are dealing with music signals, it would make sense to have the kernels the best representation of musical sounds and not represent sounds in nature like cats, dogs, etc. For the multiplicative kernels, as described in the previous section, the analogy now becomes, 

$$
\begin{aligned}
\underline{\mathcal{{F}}} \underline{f}[m] & =\sum_{n=0}^{N-1} f[n] e^{-2 \pi i m / n} \\
& =\sum_{n=0}^{N-1} f[n] \Psi_{f(input)}[m, n]
\end{aligned}
$$

where $\Psi[m, n]$ now becomes adaptive and is a function of the input itself, called $\Psi_{f(input)}[m, n]$.

\subsection{How does our work fit into the evolution} 

Starting from the representation of an audio signal, chunking it up and computing DFT and stacking the magnitude of the Fourier Transform, and taking a logarithm function is a way of computing the spectrogram, a time-frequency representation of 1-D signals. In short, a spectrogram representation is nothing but a stacked slice log of the absolute value of the Fourier Transform of the windowed signal. This can be thought of as taking a vanilla spectrogram. In the early 90s, researchers figured out that human hearing is logarithmic, so now, instead of taking for each slice the absolute value and stacking them, we can now for every slice start taking buckets onto which we are putting the energies of the slices of the STFT. These energies are then stacked instead of individual sum in the energy of bins of a DFT of a windowed signal. Different transforms are nothing but a different linear mapping from an STFT slice to that of interest, e.g., a bark scale or a mel scale. They differ by having the spacing between the center frequencies different and the bandwidths to be different. E.g., a bark scale and a mel-scale with the same number of filters, let us say 64, would differ in how each filter bank has the location of the center frequency and the bandwidth of each filter. These were present before the advent of modern deep-learning architecture. With the advent of deep learning, researchers started applying or mimicking standard signal processing pipelines via learned architectures. E.g., STFT can be interpreted as the output of a single layer of convolutional filters transformed to a vector via a linear operation. Let us say we are given a signal of length $N$, i.e., $x[n]$. We convolve this input via M convolution filters and get a signal of the same length by zero padding on either end. Let us say we learn M filters and each of which, when convolves with $x[n]$ gives an N-dimensional output (equal to the size of the input). Thus we are now left with a $MN$ matrix. However, we are interested in a $M$ dimensional vector similar to taking a DFT on a windowed signal. We then take the maximum or the average across the rows to get a single vector out of a $N$ dimensional vector. We then do similar operations that one would do on an STFT slice on this learned $M$ dimensional vector. 
\begin{figure*}[t]
  \centering
  \hspace*{8.8pt}
  \includegraphics[width=\linewidth]{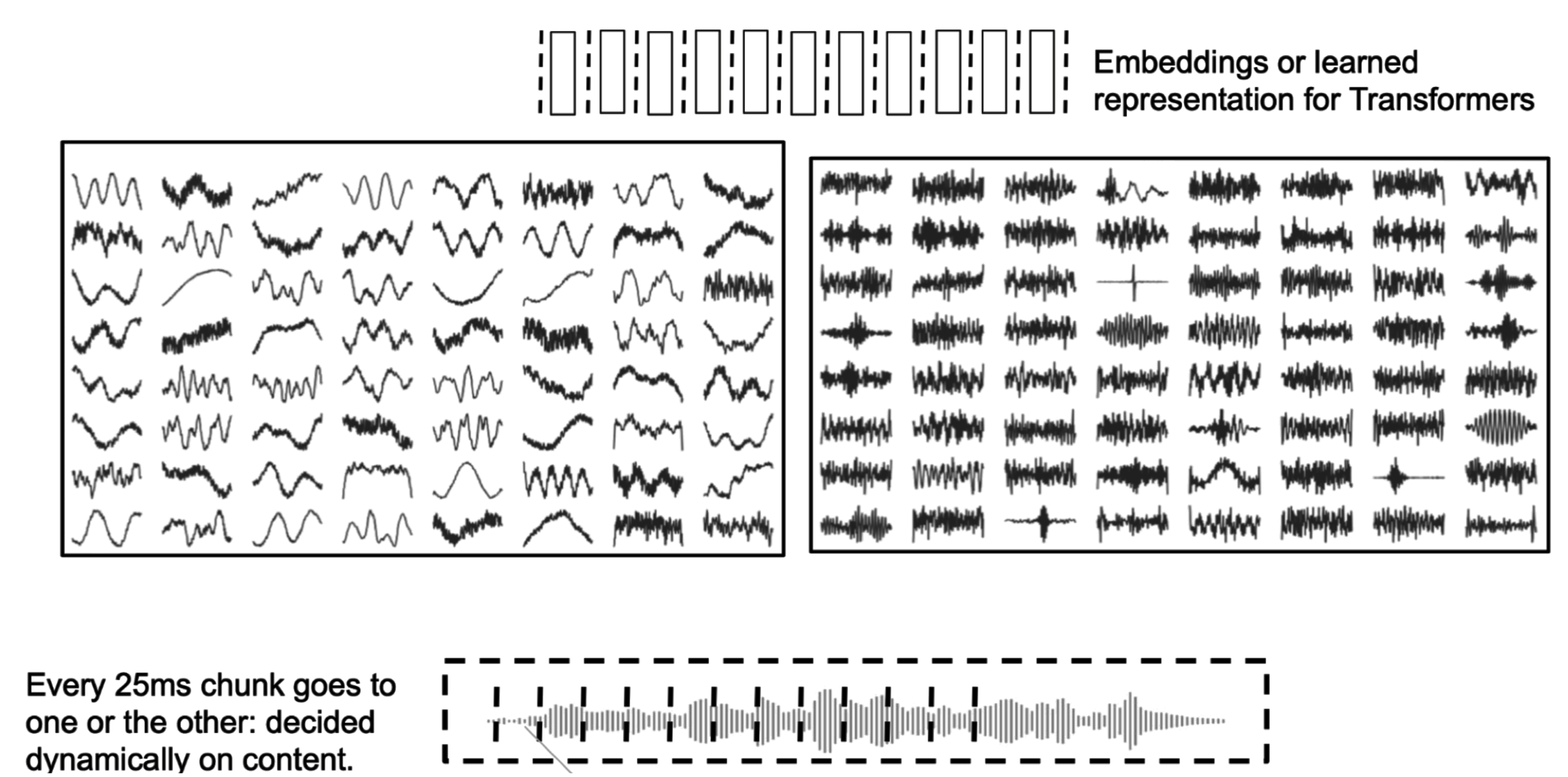}
  \caption{We plot for a content adaptive front end the output of the bank }
  \label{fig:speech_production}
\end{figure*}

\subsection{Adaptiveness of $\Psi[m, n]$  according to input} There are several ways to do this. One of the ways we follow is as follows: we learn different kernel sets and list them as $\underline{\mathcal{F}_{i}} \underline{f}[m]$, where $i$ can range from $[1, K]$. Then each of these will have learned a different $\Psi_{(i)}[m, n]$. In order to convert a list of $\Psi_{(i)}$ to that of $\Psi_{f(input)}[m, n]$, we pick one of $\Psi_{(i)}$ depending on the contents of the input signal $x[n]$. This is achieved by using a softmax function. 
As can be seen in the Figure 5, we work with two $\Psi_{(i)}[m, n]$ from the bottom left. Each of them is computing a $\underline{\mathcal{F}_{i}} \underline{f}[m]$ according to the definition of DFT as shown above with $\underline{f}[m]$ typically being 25ms waveform chunks. Now in order to make the kernel $\Psi[m, n]$ in the classic definition according to the input, we have to pick only \textit{ONE} of the $\underline{\mathcal{F}_{i}} \underline{f}[m]$. This is carried out by utilizing a softmax function. A softmax function converts the components of an $n$-vector into a probabilistic score. Thus for an $n$ vector, it is defined as, 
$$
softmax\left(x_i\right)=\frac{e^{x_i}}{\sum_{j=1}^n e^{x_j}}$$ for the $i^{th}$ component. As the name suggests, it accentuates the input data and increases
As seen from the Figure 4, we use a sparse decision router that can take into the input the same as that being passed a raw waveform for computing the transform. The output of the sparse decision route (typically a 3-layer MLP 2048 neuron architecture) is passed onto the softmax function twice. Let the output of the last layer of MLP be $x_{sr}$, then the output of the weighing functions is 
$$x_w = softmax(\alpha * softmax(x_{sr}))$$

We see that $x_w$ is chosen to be as sparse as possible, only to allow one of the normalized outputs to pass through. We differ here from the mixture of experts as depicted by \cite{jacobs1991adaptive} in the sense that we only allow one of the possible $K$ transforms instead of taking a mixture of weights of the outputs. $\alpha$ is the most significant value that can not result in an overflow. This can allow us to choose one of the possible $\underline{\mathcal{F}_{i}} \underline{f}[m]$. This is the way of making the transformed content adaptive, i.e., depending on the contents of the input signal, we make the transform adaptive. We trained our network via the cross entropy loss at the very end, as described in the preliminary section. So, the transform itself is adaptive, which means the learned kernels and how the input gets routed to the output of the kernel are all learned from scratch according to the optimization criterion of interest. In this case, to minimize the Huber loss between the target labels and the labels predicted by feeding this learned representation to a Transformer architecture.

\begin{figure*}[t]
  \centering
  \hspace*{8.8pt}
  \includegraphics[width=\linewidth]{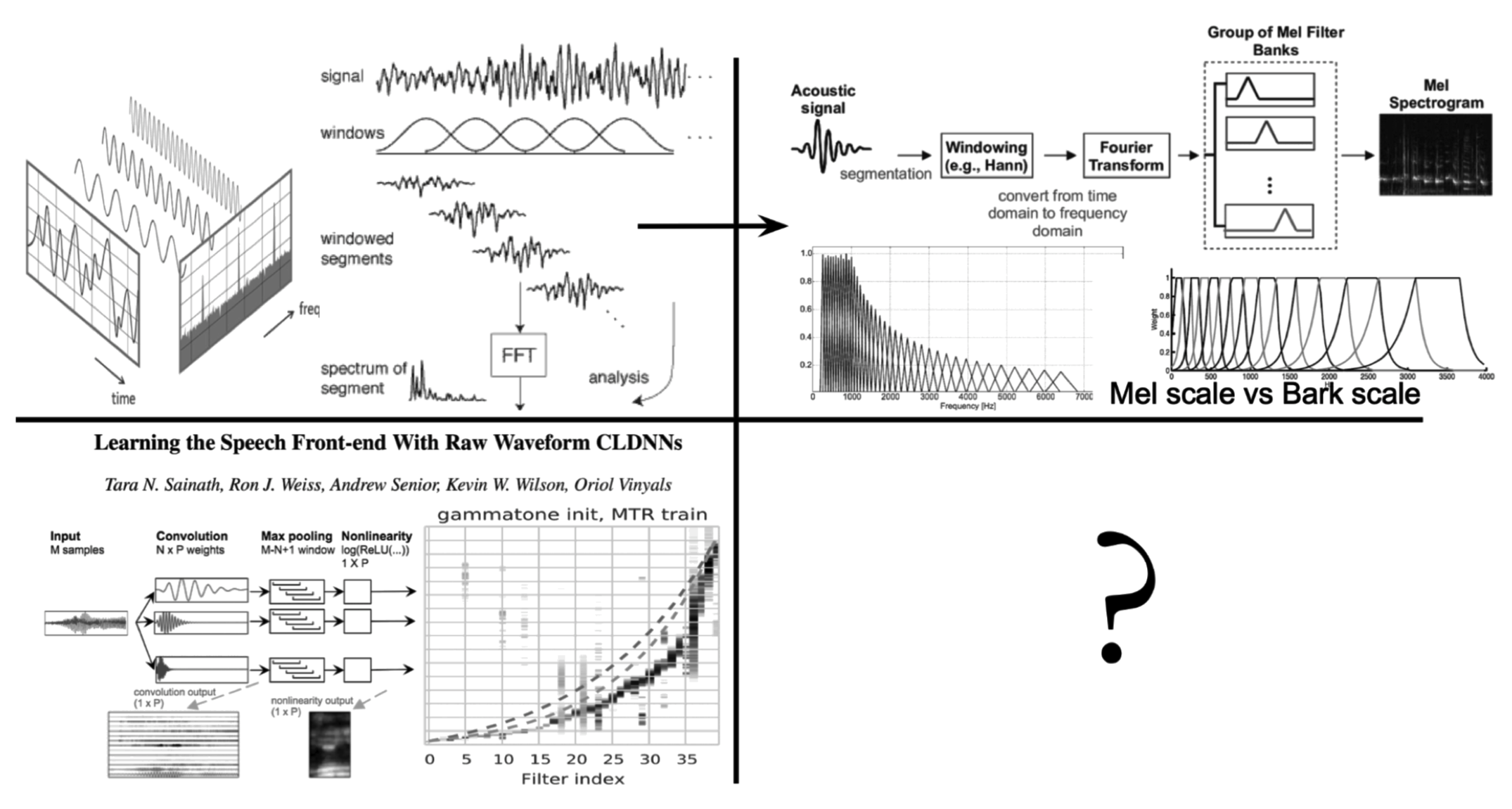}
  \caption{we show the evolution of signal representation ranging from traditional STFT to that of doing linear mapping on the slices of STFT like a mel-scale or a bark scale. With the advancements of neural architecture \cite{sainath2015learning} showed how a single layer of convolutional architecture can approximate a slice of STFT. What can the next evolution in the chain be ? We believe in our work, it can be making the transform itself content adaptive}
  \label{fig:speech_production}
\end{figure*}

\subsection{Bringing in a convolutional front end}
We will tie it down to the filter-bank interpretation of STFT. With neural architectures, this has been explored by a classic paper by \cite{sainath2015learning} by approximating a slice of a spectrogram with that of learned convolutional filters followed by the log of the absolute value of taking a maximum/average across the filter. What this achieves is mitigate a significant drawback with learning multiplicative kernels. We need not account for the phase of the individual input signals with that of the kernels, and the convolutional filters will stride along the signal to account for it. Typically zero padding is done onto the signal so that the output after doing the convolution is the same as that of the input signal. We can choose $M$ convolutional filters, which means that every input signal, $x[n]$, will give rise to $M$ signals, each having the shape of $x[n]$, with N being the length. We aim to convert these signals to a form similar to that of a slice of STFT. We average/take the maximum value of the signal across N to get a single scalar out of the convolutional filter output. Similar to a spectrogram computation, we can now take the log of the absolute value of the average/maximum, and thus this is a learnable time-frequency representation according to the task at hand. Various threads exist in the literature, e.g., SincNets \cite{ravanelli2018speaker}, that fix up the learned convolutional filter beforehand. However, this still suffers from the same drawbacks as the original Discrete Fourier or Discrete Cosine Transform formulation in that the basis functions learned from scratch will be the best "kernels" for the task at hand.

\subsection{Making the convolutional front-end content adaptive} Similar to the last section, we again use a double softmax on the output of a sparse decision router that, instead of giving a probabilistic output, tends more towards picking one out of the possible $K$ options. Now instead of choosing the output of multiplicative kernel, instead we choose the $\mathcal{F}_{i} \underline{f}[m]$ as one of the outputs of convolutions as obtained in the previous section

\subsection{What do the convolutional filters look like ?} For the best architecture that was trained, we look at how the content adaptive kernels look like for the number of experts to be 2. Thus now, depending on the contents of the input signal, we convolve it through either the set on the left or the right and get $m$ values of the transform for each of the $K=2$ experts in this case. The sparse router now takes only one of them as optimal transform. We see from the Figure 5 above that the filters exhibit quite a rich behavior and can see from the shape of the filters. The structure is also rich in terms of having a variety of shapes as opposed to a fixed sinusoidal as in STFT or Gabor wavelet functions \cite{zeghidour2021leaf}. We can see envelops, modulations, noises, in additoin to pure sinusoids. To our astonishment, a simple first-order difference function was also learned. In the first set of filter-bank, we mostly learn sinusoids or parts of sinusoids, in addition to higher frequencies and other things in the second filter-bank. As seen before, the same front end is different for various tasks, and now we find that we can learn an adaptive front end, which has been optimized for a particular task at hand, as opposed to a fixed basis function.  

\section{Conclusion} 

This report ties down several elements of the author's past research and the current research carried out with core signal processing concepts. Starting from the ground up, we have presented how classical definitions of Fourier Transforms and signal processing is connected with that of current neural architectures. This report is for people familiar with machine learning who do not know signal processing or have briefly been exposed to it. It is also for people in signal processing to see how core ideas in the field can be learned via these magical black boxes.  As a refreshing change, it does not contain a single number showing accuracies, mAP, error rates and so on. 

\section{Acknowledgement}
A part of this work is also funded by Stanford Institute for Human-Centered AI (Stanford HAI) through a generous Google Cloud Computing Grant, and we thank Google for the initiative. 

\bibliographystyle{plainnat}
\bibliography{ref}

\end{document}